# Superfluid Transition in a Chiron Gas

G. Chapline
Lawrence Livermore National Laboratory, Livermore, CA 94551

**Abstract**

Low temperature measurements of the magnetic susceptibility of LSCO suggest that the superconducting transition is associated with the disappearance of a vortex liquid. In this note we wish to draw attention to the fact that spin-orbit-like interactions in a poorly conducting layered material can lead to a new type of quantum ground state with spin polarized soliton-like charge carriers as the important quantum degree of freedom. In 2-dimensions these solitons are vortex-like, while in 3-dimensional systems they are monopole-like. In either case there is a natural mechanism for the pairing of spin up and spin down solitons, and we find that at low temperatures there is a cross-over transition as a function of carrier density between a state where the solitons are free and a condensate state where the spin up and spin down solitons in neighboring layers are paired.

The discovery that a Josephson voltage persists in $LaCuO_4$ doped with Sr (LSCO) and $Bi_2Sr_2CuO_4$ doped with La at temperatures well above the superconducting transition temperature [1] has raised the question as to how excitations similar to Abrikosov vortices can exist where there is no superconductivity. Following this discovery other anomalies were discovered in the normal sate magnetic susceptibility that pointed to the existence of vortices [2]. This mystery has deepened with the observation that in pseudo-gap region for LSCO there is a low temperature anisotropic contribution to the magnetic susceptibility that appears to be due to the presence of a "vortex liquid" which persists even for zero external magnetic field [3]. As the doping is increased at very low temperatures the vortex liquid disappears at the superconducting transition, and presumably this is a generic feature of the high $T_c$ cuprates. The layered nature of the superconducting cuprates has invited comparisons between this vortex liquid to superconducting transition and the classical Kosterlitz-Thouless (KT) transition [4]. Indeed there is now direct evidence from YBCO thin films that the superconducting transition as a function of temperature is KT-like [5]. In this note we draw attention to a mathematical model for a layered conductor which seems to provide a natural interpretation for a vortex liquid to superconductor transition. Our model, which has been discussed before [6,7], is a 3-dimensional generalization of a model for interacting anyons [8]. The charge carriers in our model are quasi-localized solitons as a result of a spin orbit-like coupling stabilizing local electric field fluctuations. In this paper we point out that at low temperatures our model has a cross-over as a function of carrier density between a state where the solitons are free and a state where the spin up and spin down solitons in neighboring layers are bound together and form a condensate.

Although one might well question whether spin orbit interactions can play an important role in the superconducting cuprates, it has been known for a long time that spin orbit effects can lead to interesting effects in transition metal oxides, especially when coupled with a deformable lattice. For example, in $Fe_2O_4$ doped with a small amount of Co spin orbit interactions in the presence of a trigonal Jahn-Teller distortion lead to a switching of the easy axis from <100> to <111>. The mechanism by which spin orbit effects in oxide materials might lead to superconductivity is illustrated by the pyrochlore $Ce_2Re_2O_7$. $Ce_2Re_2O_7$ is only known superconducting pyrochlore, and this may be related to a cubic to a tetragonal martensitic transformation that occurs near 200 K [9]. This transformation appears to be accompanied by the appearance of a "ferroelectric" metal; a possibility that was foreseen some time ago [10]. Although it is not possible to have macroscopic electric fields in a metal, short range electric field fluctuations can occur, particularly in a poorly conducting ionic material whose lattice has ferroelectric-like instabilities. The model described in the following provides a mathematical formulation of the idea that a strong spin orbit interaction acting in concert with lattice distortions can produce localized charge and spin current densities [7].

A key feature of our model is the appearance of localized spin currents. In general, quantum ground states can have non-vanishing spin currents only if inversion symmetry is broken and spin-orbit effects are important. For example, it has been predicted that spin currents will appear in semiconductors with a strong intrinsic spin orbit interactions when an external electric field is applied [11,12]. In the presence of a uniform electric field $E$ these spin currents have the form

$$j_\beta^\alpha = \sigma_s \varepsilon_{\alpha\beta\gamma} E_\gamma, \tag{1}$$

where $\sigma_s$ is the spin Hall "conductivity". In a doped semiconductor $\sigma_s$ will be proportional to the Fermi momentum, and can be comparable in magnitude to the ordinary electrical conductance. Of course, inside a good conductor there can be no macroscopic electric field due to screening, and so the spin current induced by the external field will vanish. However, in a material with a low carrier density there will be fluctuating local electric fields – which can be particularly strong if there are lattice deformations which break inversion symmetry - and hence localized spin currents might arise in such a material even in the absence of an external field.

In order to describe the effect of local in-plane electric field on charge carriers in a 2-dimensional conductor we introduce a Chern-Simons interaction which relates the effective magnetic field seen by the charge carriers to the screening length scale for the in-plane field. Neglecting spatial variations in the electric field Guass' law will be replaced by the Chern-Simons equation [13]

$$B_{eff} = -\frac{e}{\kappa}\rho,  \qquad (2)$$

where $B_{eff}$ is the magnitude of an effective magnetic field whose direction is perpendicular to the layer, $\rho$ is the charge per unit area, and $1/|\kappa|$ is an inverse length. If electric charge screening is weak $1/|\kappa|$ measures the strength of the spin orbit coupling and $B_{eff}$ can be interpreted as the effective magnetic field seen by charge carriers due to their motion in the local electric field. For carrier concentrations such that Fermi degeneracy is not important eq's (1-2) might be interpreted as the constitutive equations for a quantum fluid consisting of spin polarized carriers.

The wave function for a quantum gas of particles interacting via gauge potentials that satisfy eq.s (1-2) will satisfy a non-linear Schrodinger equation of the form:

$$i\hbar\frac{\partial \psi}{\partial t} = -\frac{1}{2m}D^2\psi + eA_0\psi - g|\psi|^2\psi,  \qquad (3)$$

where $D_\alpha = \partial_\alpha - i(e/\hbar c)A_\alpha$. The gauge fields $A_0$ and $A_\alpha$ do not satisfy Maxwell's equations, but instead are determined self-consistently from eq's (1-3) with $\sigma_s = \kappa$. The nonlinear term with coefficient $g$ represents the effect of spin orbit coupling. It was shown some time ago [13] that the time independent version of Eq. (3) in conjunction with eq's (1-2) can be solved analytically if one assumes that

$$g = \pm e^2 \hbar / mc\kappa. \tag{4}$$

The ground state contains spin polarized charge carriers with vortex-like spin currents and two units of effective magnetic flux attached to every carrier. The two signs for *g* correspond to solutions with all the carrier spins either up or down, and the effective magnetic field either parallel or anti-parallel to the z-axis, Thus these two ground state solutions form a Kramers pair.

In a material consisting of many layers one must take into account inter-layer interactions and tunneling. In ref. [6] the idea was introduced of representing the effect of having layers by replacing the scalar equation (3) with a non-abelian SU(N) Schrodinger equation:

$$i\hbar \frac{\partial \Phi}{\partial t} = -\frac{1}{2m} D^2 \Phi + e[A_0, \Phi] - g[[\Phi^*, \Phi], \Phi], \tag{5}$$

where the wave function $\Phi$ and potentials $A_0$ and $A_i$ are now N x N SU(N) matrices, and $D \equiv \nabla - i(e/\hbar c)[A,$. The number N represents the number of layers. The effective magnetic field $B_{eff} = \partial_x A_y - \partial_y A_x + [A_x, A_y]$ is now a diagonal matrix:

$$B_{eff} = -\frac{e}{\kappa}[\Phi^*, \Phi], \tag{6}$$

so the effective magnetic field seen by charge carriers can now vary from layer to layer. The in plane electric field $E_\alpha$ will also be a diagonal matrix:

$$E_\alpha = -\frac{1}{\kappa} \varepsilon_{\alpha\beta} j_\beta, \tag{7}$$

where $j_\alpha = (\hbar/2mi)([\Phi^*, D_\alpha \Phi] - [D_\alpha \Phi^*, \Phi])$ is the in-plane current. Time independent analytic solutions to eq. (5-7) can be found for any value of N if eq.(4) is satisfied. These analytic solutions represent zero energy ground states and satisfy the 2-dimensional self-duality condition $D_\alpha \Phi = \pm i\varepsilon_{\alpha\beta} D_\beta \Phi$.

In the limit N → ∞ the analytic solutions take a particularly simple form such that the effective magnetic field seen by the *j*th carrier is given by

$$B_j = \pm \frac{\hbar c}{e} \sum_k \nabla_k |X_j - X_k|, \tag{8}$$

where $X \equiv (z, u)$ is now a 3-dimensional coordinate encoding both the position $z = x+iy$ of a chiron within a layer and the height $u$ of the layer. In this solution the vortex-like carriers present in the solution for a single layer have become monopole-like objects, which were christened "chirons" in ref.6. These objects resemble polarons in that the electric charge is quasi-localized, but differ from polarons in that the charge localization also involves spin polarized currents. The ground state corresponding to (8) has zero energy and the wave function has the form [6]

$$\Psi = f(w) \prod_{k>j}^{\infty} \left[ \frac{R_{jk} + U_{jk}}{R_{jk} - U_{jk}} \right]^{1/2}, \tag{9}$$

where $R_{jk}^2 = U_{jk}^2 + 4(z_j - z_k)(\bar{z}_j - \bar{z}_k)$, $U_{jk} = u_j - u_k$, and $f$ is an entire function of the $\{\bar{z}_i\}$ in the self-dual case and $\{z_i\}$ in the anti-self-dual case. Writing the product on the rhs of eq. (9) as exp(S) defines an effective action for a chiron:

$$S = \frac{1}{2} \sum_j \ln \frac{R_j + u - u_j}{R_j - u + u_j}, \tag{10}$$

where $R_j^2 = (u - u_j)^2 + 4(z - z_j)(\bar{z} - \bar{z}_j)$. The wave function (9) resembles in some respects Laughlin's wave function for the fractional quantum Hall effect; for example moving the $z$ coordinate of a chiron around the position of a chiron in a different layer changes S by $i\pi$ [6]. However, our wave function describes a 3-dimensional system and, in contrast with the fractional quantum Hall effect, there are two distinct degenerate ground states corresponding to the self-dual and anti-self-dual solutions for eq. (5). Physically these two solutions correspond to having all the carrier spins be either up or down, and as in the 2-dimensional case form a Kramers pair.

Regarded as a classical field eq. (4) can be regarded as the Euler-Lagrange equations. for a Hamiltonian :

$$H = \int d^2z \left\{ \frac{1}{2m} |(D_x \mp iD_y)\psi|^2 - \frac{1}{2}\left(g - \frac{1}{|\kappa|}\right)\rho^2 \right\}. \tag{11}$$

It should be noted that if $g \geq e^2\hbar/mc\kappa$ increasing the charge density will decrease the potential energy, and therefore localization of the charge is favored. It might also be noted that when $g \geq e^2\hbar/mc\kappa$ applying an external field increases the energy of the soliton whose magnetic moment is parallel to the applied field, so the chiron gas is diamagnetic. In addition the classical magnetic force on the spin current associated with a spin up carrier due to the effective magnetic field of a nearby spin down carrier will point towards spin down carrier and visa versa. Thus spin orbit interactions between chirons provide a natural mechanism for forming bound pairs of spin up and spin down carriers.

Actually the effective action (10) for chirons already suggests a connection with the formation of a condensation of vortex and anti-vortex pairs in the 2-dimensional XY model. The configurations of XY spins implicated in this condensation transition have the form:

$$\Theta(z) = \sum_i m_i \operatorname{Im}\ln(z - z_i),$$

where the integer $m_i$ is the quantized circulation of the vortex (or anti-vortex if $m_i$ is negative) located at $z_i$. Now it is an elementary identity that the right hand side of (9) can be rewritten in the form

$$S = \sum_i \pm \tanh^{-1}\left(\frac{u - u_i}{R_i}\right), \tag{12}$$

which is intriguingly similar in form to a configuration of 2-dimensional XY vortices.

In order to compare the behavior of gas of self-dual and anti-self-dual chirons with the behavior of the XY model we note that phase variations in a 2-dimensional condensate can be described by a partition function of the form

$$Z = \int_0^{2\pi} D\Theta \exp[-\frac{K}{2}\int d^2\xi \frac{\partial \Theta}{\partial \xi_i}\frac{\partial \Theta}{\partial \xi_i}], \tag{13}$$

where $\Theta$ is a periodic coordinate whose period is $2\pi$ and K is a constant. It can be shown [14] that a discrete version of this theory interpolates between the low and high temperature phases of the XY model. Indeed evaluating the exponential in (13) for a configuration of vortices yields the partition function for a 2-D Coulomb gas. On the other hand substituting the chiron effective action (12) into the exponential in (13) yields:

$$Z_c = \exp-\pi K\left[\sum_{i \neq j} m_i m_j \ln \frac{R_{ij}}{|z_i - z_j|}\right], \qquad (14)$$

where the "vorticity" $m = \pm 1$ means spin up or spin down, and the sum is restricted to equal numbers of up and down spins. When the average nearest neighbor distance between chirons within a layer is less than the spacing $U_{ij}$ between a particular pair of layers, the contribution of those layers to the partition function (14) resembles the partition function of a discrete 2-D Coulomb gas, with the inter-layer spacing playing the role of the lattice spacing in the discrete Coulomb gas. An important difference though is that only chirons in different layers attract one another. In the case of a 2-D Coulomb gas the KT transition would occur at the value $K = 2/\pi$ [15], while the partition function for a trial ground state wave function which is simply a product of ground state wave functions (9) for spin up and spin down chirons corresponds to $K = 1$; i.e. just below the KT transition. Therefore while a product trial ground state wave function might be a good approximation in the dilute limit where the nearest neighbor distance between chirons is large, it doesn't take into account the formation of bound states of spin up and spin down chirons that should be relevant at low temperatures.

Evidently in our chiron model the spacing $c$ between nearest neighbor layers serves as a regulator for a KT-like transition, in that the transition in our model to a condensate state should resemble a classical KT-like phase transition when the mean separation $d$ between chirons within a plane is less than say $2c$. If we identify $d = a/\sqrt{\delta}$, where $a$ is the lattice spacing in the plane and $\delta$ is the doping, then $d = 2c$ would correspond in the case of LSCO to $\delta = .08$. The KT transition temperature $T_{KT}$ in a 2-dimensional Bose gas is $\pi\hbar^2\rho_s/2m^*$, where $m^*$ is the particle mass and $\rho_s$ is the superfluid density [15]. In our multilayer chiron model each spin up or spin down chiron is paired with an opposite spin chiron in a neighboring layer, so the superfluid density is $(\frac{1}{2})d^2$. For the values $a = 3.8$ and $m^* = 4m_e$ appropriate to LCSO the transition temperature $T_{KT}$ for $d = 2c$ would be 100°K, which is much larger than the observed transition temperature in LSCO for $\delta = .08$. However, the formation of a condensate in our model is not exactly a KT phase transition because the potentials between chirons are not simple logarithms. Instead, the condensation transition in our model will strictly speaking be a smooth cross-over, and will only resemble a phase transition for $d < 2c$. However the characteristic temperature where this cross-over takes place can be estimated in a fashion analogous to the reasoning of Kosterlitz and Thouless [16] by comparing the effective potential for spin up and spin down chirons with the 2-dimensional positional entropy of the chirons. The exponential in (14) involves all pairs of layers; however, realistically one expects that the nearest and next nearest layers are the most important. Keeping just the contribution of the nearest and next nearest layers in (14) and comparing with the entropy of the chirons assuming that they are located on a lattice with spacing $a$ leads to the following estimate for the cross-over temperature, which should be applicable for low densities of chirons such that $d > c$:

$$T_c = \frac{\pi \hbar^2}{4m^* d^2} \frac{0.5\ln(1+\frac{c^2}{d^2}) + 0.5\ln(1+\frac{(2c)^2}{d^2})}{\ln(d^2/a^2)} \;. \qquad (15)$$

A comparison of the characteristic cross-over temperature predicted by this relation for $c/a = 1.7$ with the observed transition temperatures in underdoped LSCO is shown in Fig 1. The agreement is not perfect and perhaps can be improved by including more layers. Also, it should be kept in mind that we have ignored the details of the potential between spin up and spin down chirons for separations $d < 1/|\kappa|$, and the effects of charge screening which eventually suppress the superconducting transition as the doping is increased. It should also be noted that although our predicted transition temperature is lower than the classical KT temperature, our estimate (15) for the cross-over temperature is consistent with the observation that in all cases the superconducting transition temperature in the cuprates is approximately proportional to $\rho_s / m^*$ [17].

Our analytical estimate for the characteristic temperature where pairs of spin up and spin down chiron pairs form a condensate seems to be in reasonable agreement with the observed superconducting transition temperatures in the cuprates. This in turn suggests that our picture of high temperature superconductivity as being due to the natural pairing of spin orbit localized carriers in neighboring planes is basically correct. To our knowledge this is the simplest explanation yet put forward for the basic phenomenology of the high temperature superconductivity. It is interesting to note that the soliton-like charge carriers in our model have non-trivial topological properties, and can perhaps be thought of as nano-scale analogs of the massless chiral "edge states" that occur in 2-dimensional conductors with strong spin orbit interactions; e.g. HgTe quantum wells [18]. Indeed one possibility for testing our interpretation of high temperature superconductivity might be to look for evidence of topological boundary states in high $T_c$ materials.

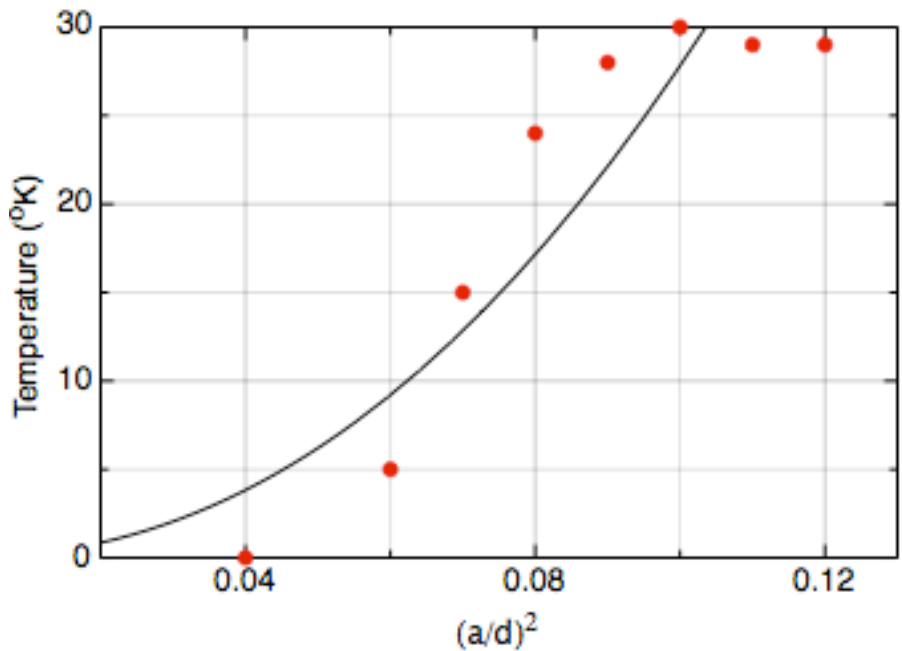

Fig 1. Comparison the expression (15) for the cross-over temperature in our chiron model (solid curve) with the observed transition temperatures in LSCO (points). The experimental transition temperatures are taken from reference 2, and it was assumed that we could identify the doping of the experimental samples with $(a/d)^2$ .

## Acknowledgements
The author wishes to thank Andre Bernevig, Mike Fluss, Bob Laughlin, Scott McCall, Kevin Moore, Gil Lonzarich, Peter Riseborough, David Santiago, James Smith, and Shou-Cheng Zhang for helpful conversations. **This work was performed under the**

## References

1. Y. Wang, Z. A. Xu, T. Kakeshita, S. Uchida, S. Ono, Y. Ando, and N. P. Ong, Phys. Rev. B 64, 224519 (2001).

2. C. Panagopoulos, M. Majoros, and A. P. Petrovic, Phys. Rev. B 69, 144508 (2004).

3. L. Lu, J. Checkelsky, S. Komiya, Y. Ando, and N. P. Ong, Nature Physics 3, 311 (2006).

4. A. Bernevig, Z. Navario, and D. Santiago, D., Phil Mag **84**, 2331 (2004).

5. I. Hetel, T. R. Lemberger, and M. Randeria, Nature Physics 3, 700 (2007).

6. G. Chapline and K. Yamagishi, Phys. Rev. Lett., **66**, 3046 (1991).

7. G. Chapline, Phil Mag **86**, 1201 (2006).

8. S. M. Girvin, A. H. MacDonald, M. Fisher, S. J.Rey, and J. Sethna, Phys. Rev. Lett., **65**, 1671 (1990).

9. H. Haawa et. al. Phys. Rev. Lett., **87**, 187001 (2001).

10. P. W. Anderson and E. I. Blount, Phys. Rev. Lett., **58**, 1252 (1965).

11. S. Murakami, N. Nagaosa, and S_C. Zhang, *Science* **301**, 1348 (2003).

12. Sinova, D. Culcer., Q. Niu, N. Sinitsyn, T. Jungwirth, and A. H. MacDonald, Phys. Rev. Lett., **92**, 126603 (2004).

13. R. Jackiw and S-Y. Pi, Phys. Rev. Lett., **64**, 3230 (1990).

14. G. Chapline and F. Klinkhamer, Mod. Phys Lett. A 11, 1063 (1989).

15. J. Jose, et. al., Phys. Rev. B, **16**, 1217 (1977).

16. J. M. Kosterlitz and D. J. Thouless, J. Phys. C 6, 1181 (1973).

17. Y. J. Uemura, et. al. Phys. Rev. Lett. 62, 2317 (1989).

18. B. A. Bernivig, T. L Hughs, and S-C. Zhang, Science 314, 1757 (2006).